\documentclass[english,twocolumn,aps,prl,superscriptaddress,showpacs,amsmath]{revtex4-1}
\usepackage[utf8]{inputenc}
\usepackage[T1]{fontenc}
\usepackage{graphicx}
\usepackage{color}
\usepackage{units}
\usepackage{amssymb}
\usepackage{esint}
\usepackage{bm}
\usepackage{amsmath} 
\usepackage{amsfonts}
\usepackage{graphicx}
\usepackage{booktabs}
\usepackage[colorlinks=true,citecolor=blue,linkcolor=blue,urlcolor=blue]{hyperref}

\begin{document}

\title{Conductance Oscillations in a Topological Insulator–Disordered Superconductor Hybrid Interface }
	
\author{Jagadis Prasad Nayak}	
\affiliation{Department of Physics and Astronomy, National Institute of Technology, Rourkela 769008, India}	

\author{Aviad Frydman}
\email{aviad.frydman@gmail.com}
\affiliation{Department of Physics, Jack and Pearl Resnick Institute and the Institute of Nanotechnology and Advanced Materials, Bar-Ilan University, Ramat-Gan 52900, Israel}

\author{Gopi Nath Daptary}
\email{daptarygn@nitrkl.ac.in}
\affiliation{Department of Physics and Astronomy, National Institute of Technology, Rourkela 769008, India}
\affiliation{Department of Physics, Jack and Pearl Resnick Institute and the Institute of Nanotechnology and Advanced Materials,	Bar-Ilan University, Ramat-Gan 52900, Israel}

\date{\today}

\begin{abstract}

We report on the observation on proximity-induced superconductivity in the topological insulator BiSbTeSe$_2$ coupled to a disordered superconductor, amorphous indium oxide (a-InO). Resistance-temperature measurements reveal superconducting signatures at low temperatures, even when InO is in an insulating state, indicating the persistence of superconducting correlations. Differential conductance spectra reveal nearly periodic oscillations at higher bias, together with a pronounced zero-bias conductance peak. Both effect disappears at high temperature, marking the critical temperature ($T^*$) of the superconducting islands in InO. These results underscore the influence of topological surface states on proximity-induced superconductivity and highlight the role of superconducting fluctuations in disordered superconductor/topological-insulator hybrid interfaces. 

\end{abstract}
\maketitle

\section{Introduction}
The study of proximity-induced superconductivity in topological insulators (TIs) has garnered significant interest in recent decades due to its potential to realize topological superconductors (TSCs) \cite{qi2011topological}. While TSCs are crucial for topological quantum computing, naturally occurring TSCs remain exceedingly rare. However, by coupling TIs with superconductors, it is possible to engineer topological superconductivity through the proximity effect, giving rise to exotic quantum states \cite{PhysRevLett.100.096407, PhysRevB.84.201105, PhysRevB.86.155431, banerjee2018signatures}.

TIs are a class of materials characterized by a bulk energy gap separating the conduction and valence bands, with topologically protected, spin-polarized surface states residing within the gap \cite{RevModPhys.82.3045, qi2011topological, xia2009observation}. These surface states, dictated by time-reversal symmetry, exhibit Dirac-like band dispersion and are robust against non-magnetic disorder. Among various TIs, bismuth-based compounds have attracted significant attention due to the clear presence of in-gap surface states observed in angle-resolved photoemission spectroscopy (ARPES) measurements \cite{lohani2017band, PhysRevB.88.121404}. While Bi$_2$(Se,Te)$_3$ materials have been widely explored \cite{PhysRevB.87.121111, chen2009experimental, PhysRevLett.104.016401, PhysRevB.88.041404}, intrinsic defects such as selenium and tellurium vacancies complicate the distinction between bulk and surface transport. In this context, BiSbTeSe$_2$ (BSTS) has emerged as a strong TI that enables the separation of surface and bulk states through transport measurements \cite {xu2014observation, biswas2019resistance}, making it an ideal platform for investigating proximity-induced superconductivity.

Previous studies have demonstrated that TI superconductor heterostructures can host a variety of unconventional proximity-induced phenomena \cite{PhysRevB.84.165120, PhysRevLett.109.056803, veldhorst2012josephson, zareapour2012proximity, PhysRevB.85.104508, banerjee2018signatures, PhysRevB.96.075107, stolyarov2021superconducting, trang2020conversion}. These studies predominantly use conventional s-wave superconductors with high carrier densities ($n$) compared to TIs, often requiring heavy doping of the TI region to facilitate the proximity effect. However, coupling a TI with a low-density, disordered superconductor can give rise to unique quantum state, as their comparable carrier densities may lead to novel superconducting correlations and quantum states \cite{PhysRevB.105.L100507, PhysRevMaterials.8.084802}.

In this report, we explore proximity-induced superconductivity in a BSTS flake coupled to a low-density, disordered superconductor, amorphous indium oxide (a-InO), under varying back-gate voltages ($V_g$) and temperatures ($T$). Resistance–temperature measurements show a distinct drop in resistance, indicating the onset of superconductivity at low temperatures, even though the InO layer itself remains in an insulating state. Differential conductance measurements exhibit a pronounced zero-bias conductance peak accompanied by nearly periodic conductance oscillations above the superconducting gap. Our results highlight the interplay between topological surface states and superconducting fluctuations, offering new insights into the nature of proximity-induced superconductivity in TI/disordered superconductor interfaces.

 \section{Experimental details}
We used BSTS flakes exfoliated from the bulk crystal and then transfer to SiO$_2$/Si substrate ( bulk crystal were purchsed from hQ graphene company). The samples were patterned into a Hall bar geometry using standard electron beam lithography and contacted with Ti/Au leads (5 nm/30 nm). A second lithography step was performed for the deposition of a-InO films via electron-gun evaporation. During the deposition, dry O$_2$ gas was introduced into the chamber at a controlled partial pressure, enabling precise tuning of the system across the superconductor-insulator transition (SIT). To enable electrostatic gating, a thin Au layer was evaporated onto the backside of the Si wafer, forming one plate of a parallel-plate capacitor, with BSTS acting as the second plate and SiO$_2$ serving as the dielectric. By varying the gate voltage applied to the Si wafer, we modulated the carrier density in the BSTS device. The device structure, including the BSTS flake and Ti/Au leads, is illustrated in Fig. \ref{fig2} (b) (lower inset). Electrical transport measurements were carried out using a Keithley sourcemeter and a standard lock-in detection technique (SR830). All measurements were performed in a wet helium cryostat at temperatures down to 1.5 K.

\begin{figure}
\includegraphics[width=8.5cm,height=!]{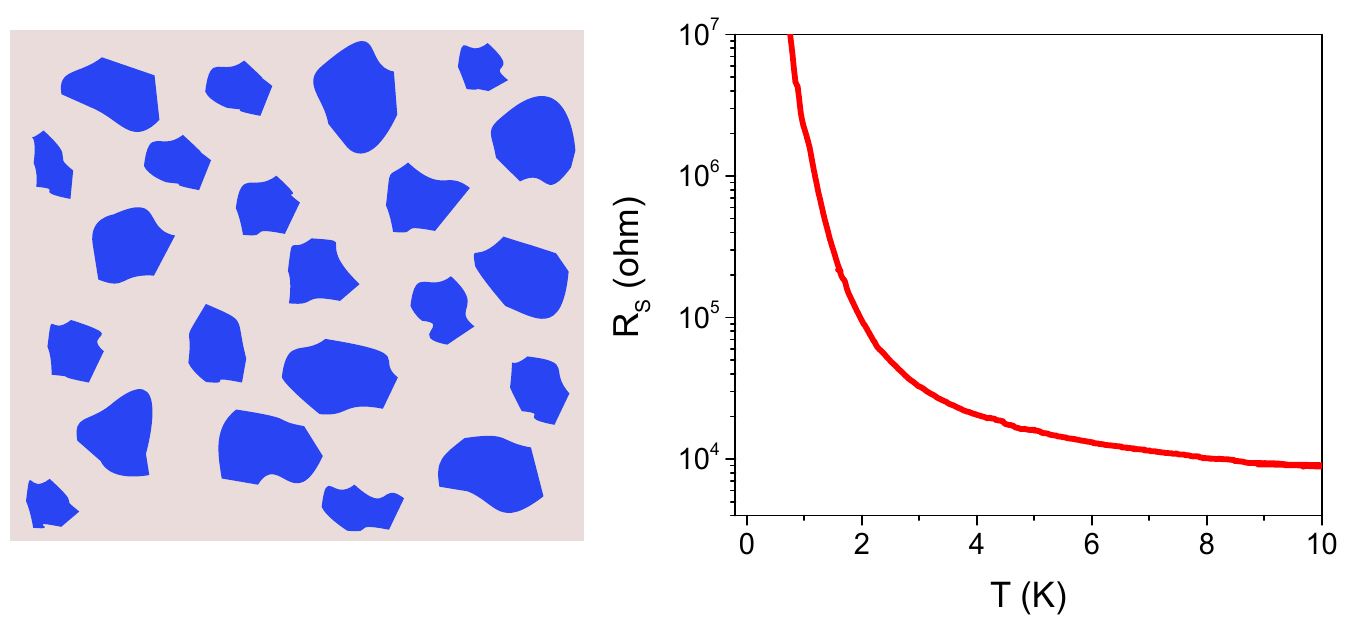}
\caption{\label{fig1}
Sketches of the superconducting islands (left) and plot of resistance-temperature of InO sample when it is in insulating state (right). 
}
\end{figure}


In our efforts to induce low-density disordered superconductivity in BSTS, we utilized 30 nm thick films of a-InO as proximity superconductors. Despite the morphological uniformity of these films, previous studies have shown that InO exhibits emergent granularity, characterized by superconducting puddles embedded within an insulating matrix~\cite{kowal1994disorder, kowal2008scale, bouadim2011single}. Experimental investigations have further revealed the presence of superconducting vortices and a finite energy gap even in the globally insulating phase of InO~\cite{sacepe2011localization, poran2011disorder, sherman2012measurement, kopnov2012little, roy2018quantum}. These findings suggest that the loss of global phase coherence in InO does not necessarily lead to the closure of the superconducting pairing gap ($\Delta$), as isolated superconducting islands remain. This intriguing state, where local electronic pairing coexists with an insulating background, is often referred to as a "Bosonic insulator."
The disorder inherent in InO films distinguishes between the temperature $T^*$ at which superconducting pairing first emerges within individual grains - leading to a soft gap in the local density of states - and the global resistive critical temperature $T_C$, where phase coherence across the sample establishes finite superfluid density \cite{RevModPhys.91.011002}.

For our experiments, InO films were deposited onto BSTS flakes using electron beam evaporation (see Fig. \ref{fig2} (b), lower inset). The carrier density ($n$) of the disordered superconductor, controlled by adjusting the oxygen partial pressure during deposition (2.5 $\times$ 10$^{-5}$ Torr), was determined to be in the range of $10^{19}-10^{20}$ cm$^{-3}$~\cite{ovadyahu1986some}. This carrier density is significantly lower - by several orders of magnitude - compared to conventional metals. By decreasing the oxygen partial pressure, $n$ increases, leading to a reduction in the sheet resistance ($R_s$) and eventually driving the InO films through a superconductor-insulator transition (SIT), which can be tuned via carrier concentration \cite{PhysRevMaterials.8.084802}.
In our study, the InO films remained in the insulating regime. Figure \ref{fig1} presents a plot of $R_s$ as a function of $T$ for bare InO, illustrating its insulating behavior alongside schematic representations of its inherent superconducting granularity. In the insulating phase, the superconducting islands are sparse and decoupled, resulting in only localized superconductivity (Fig. \ref{fig1}, left panel).

\section{results and Discussion}

\begin{figure} 
\includegraphics[width=8.5cm,height=!]{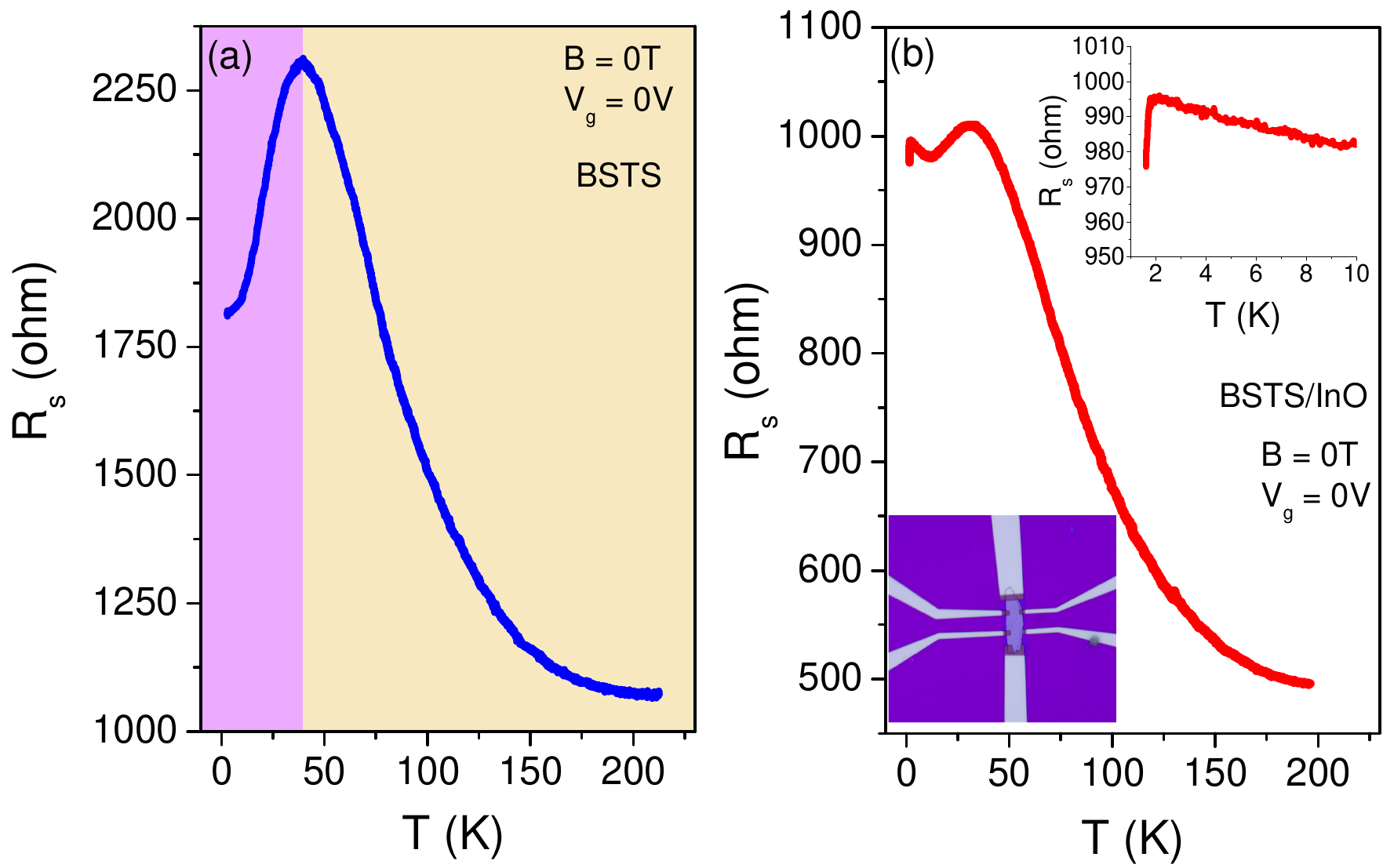}
\caption{\label{fig2}
(a) The sheet resistance, $R_s$, is plotted as a function of temperature, $T$, for a BSTS flake with a thickness of 150 nm  at $B=0$ T and $V_g=0$ V. The pink and yellow shaded regions indicate surface-dominated and bulk-dominated transport regimes, respectively. 
(b) $R_s$ vs $T$ of BSTS/InO bilayer at $B = 0$ T and $V_g = 0$ V. A resistance drop is observed at low temperatures, as highlighted in the zoomed - in view of the low-temperature data in the upper inset. The lower inset presents an optical image of the BSTS/InO bilayer, including electrical contacts.}
\end{figure}

We start the results with bare BSTS flake of thickness 150 nm. Figure \ref{fig2}(a) presents a plot showing the sheet resistance, $R_s$, versus temperature, $T$ for the BSTS flake under conditions of $B=0$ T and $V_g=0$ V. At temperatures above 40 K, $R_s$ decreases as $T$ increases. This behavior aligns with previous reports on BSTS samples and is typically attributed to the bulk contribution of the TI. Notably, below 40 K, the resistance begins to decrease further with decreasing temperature. This decrease is likely due to the emergence of resistance from parallel surface channels contributing to the overall bulk resistance. This phenomenon of decreasing $R_s$ at low temperatures has been observed in high-quality, exfoliated BSTS flakes of similar thickness and has been interpreted as a metallic contribution arising from the topological surface states \cite{xu2014observation, biswas2019resistance}.


As mentioned, insulating InO films exhibit a unique structure with emergent superconducting islands dispersed within an insulating matrix. These superconducting puddles possess a higher electron density compared to the surrounding insulating background. When coupled with a BSTS flake, this configuration creates a distinctive scenario. Through the proximity effect, the BSTS beneath these superconducting islands develops regions with a non-vanishing superconducting gap just below the islands. This induces superconducting correlations within the surface of the BSTS at low temperatures, leading to a noticeable drop in resistance, as depicted in Fig. \ref{fig2}(b). For clarity, we present a zoomed-in view of the low-temperature region, revealing an onset superconducting transition temperature of $T_C \sim 2$ K.  At higher temperatures, the $R_s$ vs $T$ curve exhibits behavior similar to that of a bare BSTS flake: insulating behavior at high temperatures (attributed to the bulk) and metallic behavior at low temperatures (attributed to the surface). The resistance drop observed at low temperatures  primarily arises from superconducting fluctuations due to the presence of InO. In the temperature range
$2<T<12$ K, the decrease in resistance with increasing temperature likely stems from the non-monotonic transport behavior of the InO film. These observations underscore the complex interplay between the BSTS surface and the proximity-induced superconductivity from the InO film. 

\begin{figure} 
\includegraphics[width=7cm,height=!]{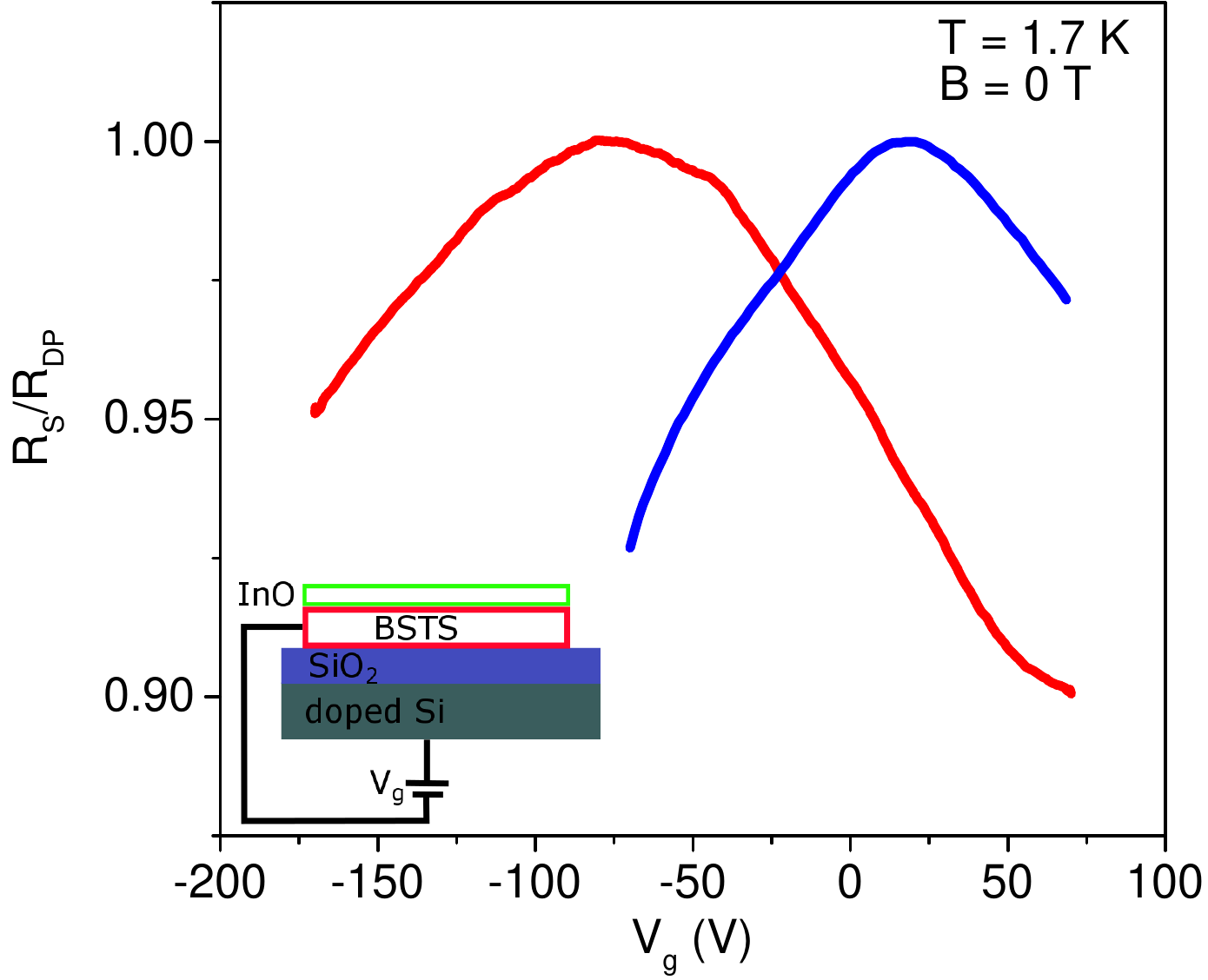}
\caption{\label{fig3}
(a) Sheet Resistance, $R_s$ of bare BSTS (blue solid line) and bilayer BSTS/InO (red solid line)  as a function of back gate voltage, $V_g$ at $B=0$ T and $T=1.7$ K. In the inset (lower panel), a typical device structure along with gate connection is shown. 
}
\end{figure}

To further explore the influence of superconducting fluctuations on surface charge carriers, we performed resistance measurements as a function of back-gate voltage, $V_g$ at $T=1.7$ K. The carrier density of the bilayer device can be tuned using $V_g$ due to the 2D nature of BSTS surface states, as the carrier density of InO is relatively high. The schematic of the back-gated device is depicted in Fig. \ref{fig3} (lower panel). Figure \ref{fig3} shows the sheet resistance $R_s$ of the bare BSTS (blue solid line) and bilayer device (red solid line) as a function of $V_g$ at $B=0$ T and $T=1.7$ K. In the bare BSTS device, $R_s$ varies by approximately 10\% with changing $V_g$, peaking at $V_g = 16$ V. This ambipolar transport behavior suggests the coexistence of electron and hole carriers, with the charge neutrality point (Dirac point) at $V_{DP}=16$ V.
Similar ambipolar characteristics, indicative of 2D surface-dominated transport, have been reported in comparable systems \cite{xu2014observation}. Upon covering the BSTS with an InO film, the Dirac point shifts to negative gate voltages, indicating electron doping in the BSTS surface layer. This shift further confirms the emergence of proximity-induced superconducting correlations on the surface state of TI.

\begin{figure} 
\includegraphics[width=8.2cm,height=!]{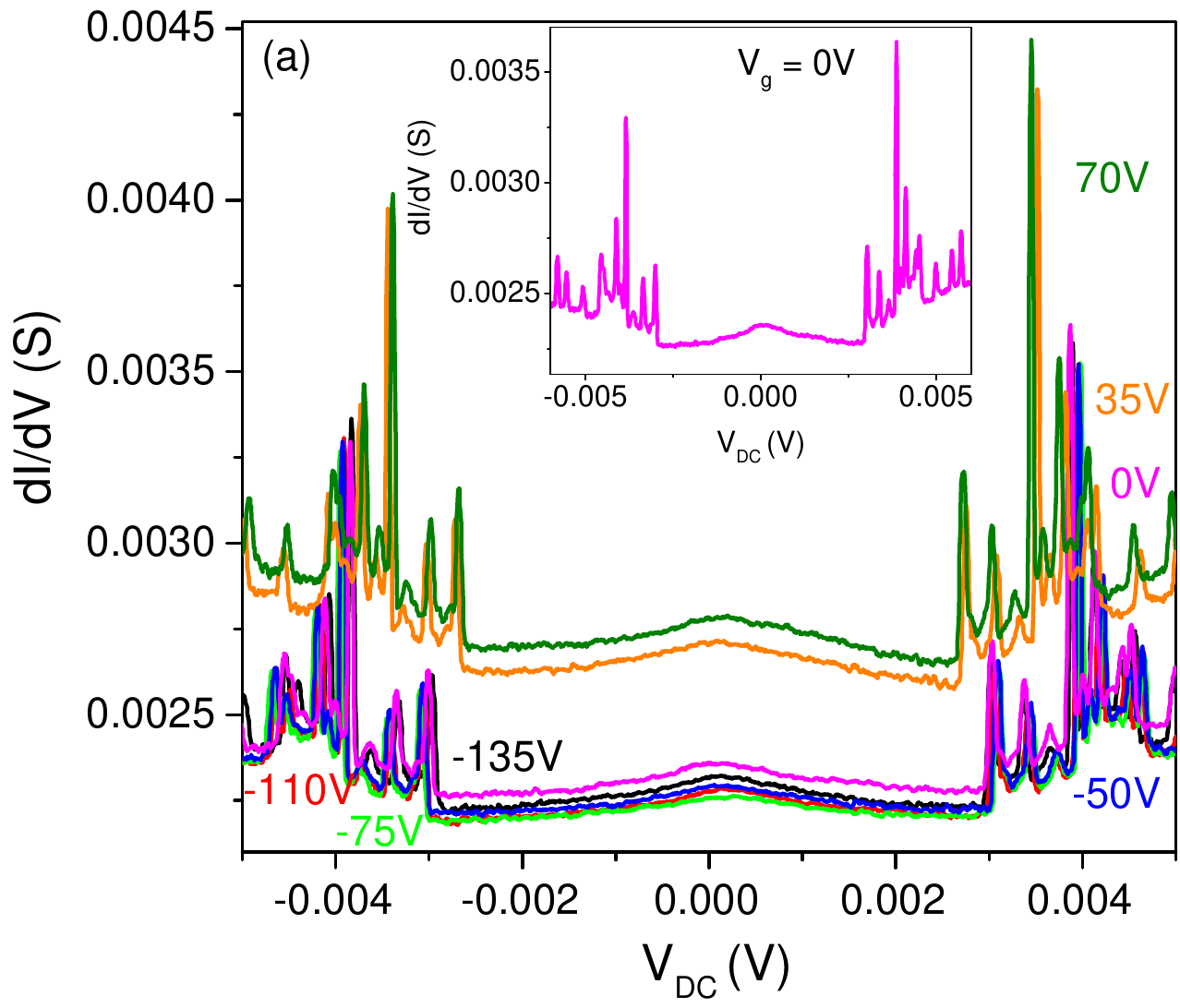}
\caption{\label{fig4}
Differential conductance, $dI/dV$, of the BSTS/InO bilayer is plotted as a function of bias voltage, $V_{DC}$, at various gate voltages, $V_g$, for $T = 1.7$ K and $B = 0$ T. Note that a conductance peak is observed at $V_{DC} = 0$ V, indicative of enhanced Andreev reflection. Additionally, multiple conductance peaks appear at higher $V_{DC}$. For clarity, the data for $V_g = 0$ V is shown in the inset.}
\end{figure}

Figure \ref{fig4} presents differential conductance ($dI/dV$) of the bilayer as a function of bias voltage (V$_{DC}$) at various $V_g$ at $T=1.7$ K. A prominent zero-bias conductance (ZBC) peak is observed, suggesting enhanced Andreev reflection at the interface between the TI and the superconducting islands within InO layer. Andreev reflection occurs when an electron incident from the normal region is retroreflected as a hole at the interface, forming a Cooper pair in the superconductor. This process is highly sensitive to the quality of the interface, and a pronounced ZBC peak typically indicates strong coupling between the superconducting and topological states. 

Oscillatory features observed in the $dI/dV$ of superconductor junctions arise from the phase-coherent interference of quasiparticles (electron- and hole-like excitations) undergoing multiple reflections at the interfaces. These oscillations serve as a powerful probe for key physical parameters, including the coherence length ($\xi_S$), Fermi velocity ($v_F$), interface transparency, and the geometry of the junction \cite{tomasch1965geometrical,visani2012equal}. The formation of these conductance oscillations is contingent upon the cavity length, $L$, representing the layer thickness or effective path, being comparable to the phase coherence length ($L_{\phi}$) or the superconducting coherence length ($\xi_S$), which establishes standing-wave conditions.

Quasiparticle interference in superconductor-based heterostructures manifests through three principal oscillatory phenomena-Multiple Andreev Reflection (MAR), Tomasch, and McMillan–Rowell (MR) oscillations-each governed by distinct confinement conditions and transport regimes. MAR arises from successive Andreev reflections in Superconductor–Normal–Superconductor (SNS) junctions, where quasiparticles traverse the junction multiple times before losing phase coherence \cite{octavio1983subharmonic}. The subharmonic gap structures appear when $eV = \frac{2\Delta}{n}$,
with integer $n$ and superconducting gap $2\Delta$. These symmetric subgap features ($|V| \le 2\Delta$) are characteristic of coherent multiple charge transfer through transparent interfaces. However, such MAR features are absent within the gap.
${Tomasch}$ oscillations originate from interference between electron-like and hole-like quasiparticles within the superconducting layer of thickness $L_S$ \cite{tomasch1965geometrical}. The resonance condition is
\begin{equation}
  eV_n = \sqrt{\Delta^2 + \left(\frac{n\pi\hbar v_F}{2L_S}\right)^2}  
\end{equation}
where $V_F$ is the Fermi velocity. This leads to asymmetric oscillations observed predominantly above the gap ($|V| > 2\Delta$). The voltage spacing follows $\Delta V_T \approx \frac{\pi \hbar v_F}{2eL_S}$.
${McMillan-Rowell}$ oscillations arise from the interference between electron and hole trajectories within the normal layer, bounded by the NS interface and its outer surface \cite{mcmillan1966theory}. The voltage spacing reflects the normal-layer thickness $L_N$ is given by $\Delta V_{MR} = \frac{h v_{F,N}}{4eL_N}$.
These oscillations usually appear within or near the superconducting gap and indicate coherent quasiparticle reflection inside the normal region. When $L_{\mathrm{eff}} \!\sim\! L_S$, Tomasch-type oscillations dominate; when $L_{\mathrm{eff}} \!\sim\! L_N$, MR oscillations prevail. 

The oscillatory features observed in the differential conductance $dI/dV$ likely indicate a hybrid regime exhibiting characteristics of both McMillan–Rowell and Tomasch oscillations. We quantitatively estimated the effective normal and superconducting path lengths, $L_N$ and $L_S$, from the measured oscillation spacing. Notably, the oscillation spacing is quasi-periodic, varying from approximately 1.6 mV to 1.2 mV as the gate voltage $V_g$ is tuned from –135 V to 70 V. Using the Fermi velocity $V_F \approx 3 \times 10^5$m/s reported from ARPES measurements \cite {arakane2012tunable, neupane2012topological}, we obtain $L_N \approx 138-250$ nm and $L_S \approx 190-260$ nm
over the same gate-voltage range. The extracted value of $L_N$ is comparable to the BSTS thickness ($\sim150$ nm), suggesting that the McMillan–Rowell mechanism plays a more dominant role than Tomasch oscillations. In the absence of a comprehensive theoretical model that fully accounts for these observations, a detailed explanation of the conductance oscillations is left for future investigation.


 
\begin{figure} 
	\includegraphics[width=7cm,height=!]{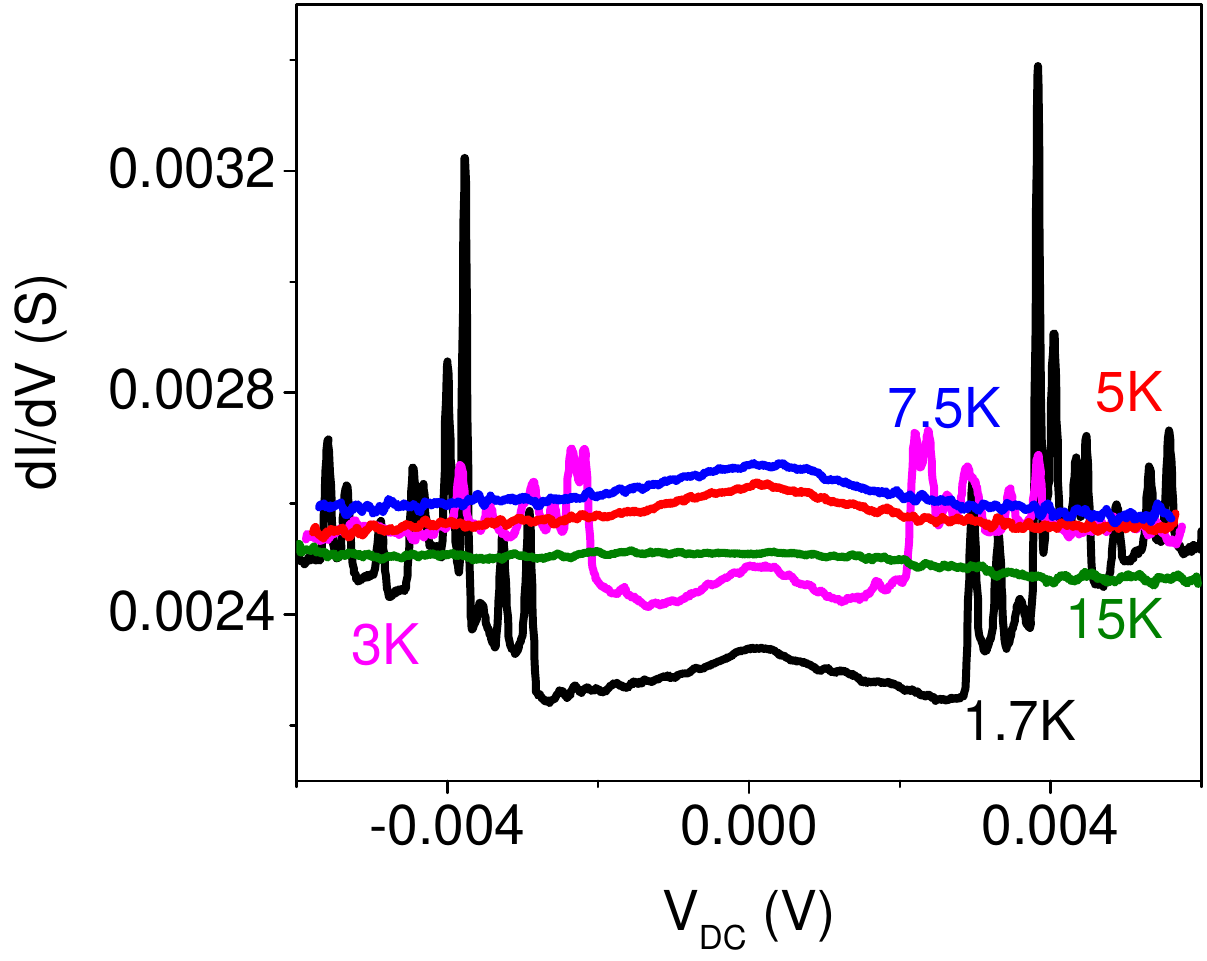}
	\caption{\label{fig5}
		Differential conductance, $dI/dV$ of BSTS/InO bilayer as a function of bias voltages, $V_{DC}$ at different temperature for $V_g=-170$ V and $B=0$ T. 
}
\end{figure}

To further investigate the temperature dependence of conductance oscillations, we measured $dI/dV$ as a function of V$_{DC}$ at a $V_g = -170$ V under zero magnetic field. As shown in Fig. \ref{fig5}, the ZBC as well as oscillations decrease progressively with increasing temperature and eventually ZBC vanish at $T=15$ K. This means that superconductivity in BSTS persists up to approximately 10 K, even though the global superconducting transition temperature, $T_C$ of InO is expected to be significantly lower. This observation is consistent with scanning tunneling microscopy (STM) studies on InO films with a global $T_C$ of approximately 3 K, where a finite local superconducting gap ($\Delta$) was detected up to $T \sim 6.5$ K \cite{sacepe2011localization}. Theoretical studies further predict that in insulating regimes, the local superconducting gap can increase with disorder \cite{bouadim2011single}. The true pairing critical temperature ($T^*$) of the superconducting islands within InO remains undetermined; however, our results suggest that $T^*$ could reach as high as 10 K, even in the context of proximity-induced superconductivity in topological insulators.

\section{Conclusion}

 In conclusion, we report the observation of proximity-induced superconductivity in an exfoliated, bulk-insulating 3D topological insulator coupled to a disordered InO superconductor. Superconductivity emerges when transport is dominated by the 2D topological surface states. Additionally, we observe a zero-bias conductance peak and conductance oscillation features across all gate voltages, which vanish above 10 K - significantly higher than the global $T_C$ of InO. These findings provide important insights into the interplay between topological surface states and proximity-induced superconductivity, paving the way for further investigations into hybrid topological insulator-superconductor systems.


\vspace{0.5cm}

 We are grateful for illuminating discussions with Udit Khanna. G.N.D. and A.F were supported by the Israel National Fund, ISF Grant No. 1499/21. G.N.D. acknowledges support from the ANRF IRG (Grant No. ANRF/IRG/2024/000714/PS) and the ANRF PMECRG (Grant No. ANRF/ECRG/2024/001810/PMS), Government of India. G.N.D. also gratefully acknowledges the start-up research grant provided by NIT Rourkela, India.

%

\end{document}